 \documentclass[prb,twocolumn,showpacs,amsmath,amssymb,superscriptaddress]{revtex4}
 \usepackage{graphicx}
\usepackage{dcolumn}
\usepackage[sort&compress]{natbib}
\usepackage{subfigure}
\usepackage{ifpdf}
\usepackage{bm}
\ifpdf
\usepackage[pdftex,
        colorlinks=true,
        pdftitle={},
        pdfauthor={S. Salem-Sugui Jr.},
        pdfsubject={fluctuation in iron-pnictides},
        pdfkeywords={IF, UFRJ, Rio de Janeiro Brazil},
        baseurl={http://www.if.ufrj.br},
        pdfpagemode=UseNone,
        bookmarksopen=true
        pdfoutput=1
        ]{hyperref}
\fi

\begin{document}
\title{Plastic pinning replaces collective pinning as the second magnetization peak disappears in the pnictide superconductor Ba-KFe$_2$As$_2$}
\author{Shyam Sundar}
\affiliation{Instituto de Fisica, Universidade Federal do Rio de Janeiro,
21941-972 Rio de Janeiro, RJ, Brazil}
\author{S. Salem-Sugui Jr.}
\affiliation{Instituto de Fisica, Universidade Federal do Rio de Janeiro,
21941-972 Rio de Janeiro, RJ, Brazil}
\author{H. S. Amorim}
\affiliation{Instituto de Fisica, Universidade Federal do Rio de Janeiro,
21941-972 Rio de Janeiro, RJ, Brazil}
\author{Hai-Hu Wen}
\affiliation{National Laboratory of Solid State Microstructures and Department of Physics, Innovative Center for Advanced Microstructures, Nanjing University, Nanjing 210093, China}
\author{K. A. Yates}
\affiliation{The Blackett Laboratory, Physics Department, Imperial College London, London SW7 2AZ, United Kingdom}
\author{L.F. Cohen}
\affiliation{The Blackett Laboratory, Physics Department, Imperial College London, London SW7 2AZ, United Kingdom}
\author{L. Ghivelder}
\affiliation{Instituto de Fisica, Universidade Federal do Rio de Janeiro,
21941-972 Rio de Janeiro, RJ, Brazil}

\date{\today}
\begin{abstract}
We report a detailed study of isofield magnetic relaxation and isothermal magnetization measurements with $H$$\parallel$c on an underdoped Ba$_{0.75}$K$_{0.25}$Fe$_2$As$_2$ pnictide single crystal, with superconducting transition temperature $T_c$ = 28 K. The second magnetization peak (SMP) has been observed at temperatures below $T_c$/2 and vanished at higher temperatures. The observed behaviour of the SMP has been studied by measuring the magnetic field dependence of relaxation rate, $R(H)$ and by performing the Maley's analysis. The results suggest that the crossover from collective to plastic pinning observed in the SMP disappears above 12 K with plastic pinning replacing collective pinning. An interesting $H$-$T$ phase diagram is obtained. The critical current density ($J_c$) was estimated using Bean's model and found to be $\sim$  $3.4 \times 10^9$ A/m$^2$ at 10 K in the SMP region, which is comparable to an optimally doped Ba-KFe$_2$As$_2$ superconductor and may be exploited for potential technological applications. The pinning mechanism is found to be unconventional and does not follow the usual $\delta l$ and $\delta T_c$ pinning models, which suggest the intrinsic nature of pinning in the compound. 
\end{abstract} 
\maketitle 
\section{Introduction}
The study of vortex-dynamics in type-II superconductors and especially in high temperature superconductors (HTSC) including the iron-pnictides is of great interest due to their high superconducting transition temperature ($T_c$) and the potential for future technological applications.\cite{bla94, yes96} Vortex-dynamics in iron-based superconductors (IBS) \cite{yoi08} has been the subject of intensive research interest because of their moderately high $T_c$, \cite{an08} high upper critical field ($H_{c2}$), \cite{sen08, jar08} small anisotropy \cite{yua09, alt08} and strong inter-grain connectivity, \cite{tak11b, wei12} which also makes them suitable for applications. \cite{asw10, wei16, pur15, tos15, mis16} Among the different phases existing in the vortex-phase diagram of type-II superconductors, one of the most interesting and possibly the most studied is the second magnetization peak (SMP), which is present in isothermal M(H) curves and associated to a peak in the critical current. During the past few years, numerous studies have been performed to understand the origin of SMP in different families of iron-pnictides and there is still an ongoing research with newly prepared materials. \cite{wei16, ge13, ahm14} In the literature, it has been found that the mechanism responsible for the SMP appears to be system dependent, with explanations including a crossover from elastic to plastic, \cite{wei16, said10} order-disorder transition, \cite{miu12, joh14} vortex-lattice phase transitions, \cite{kop10, pra11} and it is even still unclear for few compounds. \cite{said11, said13} For most of the systems exhibiting SMP, the associated line in the phase diagram corresponding to the peak field ($H_p$) extends from very low temperatures (with an exception in the case of Bi-2212 \cite{tam93, yes94}) up to the temperatures close to the irreversibility line, where, in the case of pnictides, the SMP disappears only as $T$ approaches $T_{irr}$. Despite the origin of SMP being known for many compounds, a fundamental question still remains: why is this feature absent in some samples? A recent study by Song et. al. \cite{son16} probed the doping dependence of the superconducting properties in Ba$_{1-x}$K$_{x}$Fe$_2$As$_2$. Surprisingly, it was found that the critical current density is higher in the $x$ = 0.30 underdoped compound, and not in the optimally doped $x$ = 0.40 composition, which is most commonly employed for application. \cite{gao15}

This result motivated us to perform a thorough study of vortex dynamics in an underdoped Ba$_{1-x}$K$_{x}$Fe$_2$As$_2$ compound. We employed a hole doped Ba$_{0.75}$K$_{0.25}$Fe$_2$As$_2$ single crystal with $T_c$~=~28~K and measured  isothermal $M(H)$ and magnetic relaxation $M(t)$ with the field parallel to the c axis ($H$$\parallel$c) of the sample. We observed that the SMP exists only up to approximately $T_c$/2 and disappear at higher temperatures. To the best of our knowledge, this is the only system, where the SMP doesn't lie in the whole temperature range below $T_c$. This unusual and interesting phenomenon, the disappearance of the SMP at higher temperatures, allowed us to study in detail how the vortex dynamics evolves as the SMP fades out. In principle, the pinning mechanisms above and below temperature $T_c$/2 should be of different nature. In order to identify the relevant pinning mechanism across the phase diagram of the compound, we performed detailed measurements of magnetic relaxation below the SMP onset ($H_{on}$), above the SMP ($H_p$) and in the region between $H_{on}$ and $H_p$ for various isothermal $M(H)$ measurements. To address the question of why the SMP in the present sample exists up to $T_c/2$ and disappears at higher temperatures, we compared the magnetic relaxation measured above and below $T_c/2$. At selected isothermal $M(H)$ curves, magnetic relaxation data were taken for magnetic field values ranging from just above $H_{c1}$ up to field values close to the irreversibility point $H_{irr}$. We estimated the activation energy using magnetic relaxation data and studied the vortex-dynamics in different $H$ and $T$ ranges of interest. We also measured the $T_c$ distribution over the sample surface and concluded that the sample inhomogeneity does not play a significant role in the pinning distribution. Our analysis showed that the disappearance of the SMP above $T_c$/2 is due to plastic pinning replacing collective pinning. We also obtained the $J_c$ values using isothermal $M(H)$ measurements and compared them with other underdoped and overdoped Ba$_{1-x}$K$_{x}$Fe$_2$As$_2$ superconductors. \cite{son16} 

\section{Experimental} 
In the present work, we studied a single crystal of Ba$_{1-x}$K$_x$Fe$_2$As$_2$ with $x$~=~0.25, in the underdoped region. The crystal was prepared using the flux method. \cite{luo08} Magnetic measurements were performed with a vibrating sample magnetometer (VSM, Quantum Design, USA). The magnetic field dependence of the magnetization, $M(H)$ and the magnetic relaxation, $M(t)$ were  measured with $H$$\parallel$c axis in zero field cooled (zfc) mode. Isothermal 
magnetization, $M(H)$ was obtained at different  temperatures, ranging from 2 K to $T_c$ with $H$ varying from 0 to 9 T. Relaxation data, $M(t)$ were taken over a period of approximately 2 hours in the increasing field branch of selected isothermal $M(H)$ curves, and for fixed magnetic fields at various different temperatures. The sample quality ($T_c$ distribution) was investigated using a scanning Hall probe, with a 5 $\mu$m $\times$ 5 $\mu$m active area Hall sensor (1 $\mu$m thick InSb epilayer on undoped GaAs substrate). \cite{per02} To map the $T_c$ values over the sample, the Hall voltage profile was recorded while scanning the sample surface. The resolution of the recorded Hall image is 256 pixels $\times$ 256 pixels. A 4 T split coil superconducting magnet and a continuous flow helium cryostat (Oxford Instruments Ltd.) were used to perform the measurements. The data was collected with an applied field of 0.01 T parallel to the c axis. X-ray diffraction analysis was performed using the transmission La{\"u}e method. A Philips X-ray generator, model PW1024 was used, with a molybdenum anode X-ray tube (15 mA and 30 kV) and a 0.8 mm collimator, to generate the x-rays. The lauegrams were obtained with the single crystal at 35.0 mm from the film and 4 hours of irradiation. Lauegrams simulations were performed using the OrientExpress software, version 3.4. \cite{lau}
\section{Results and discussion}
\begin{figure}
\centering
\includegraphics[height=11cm]{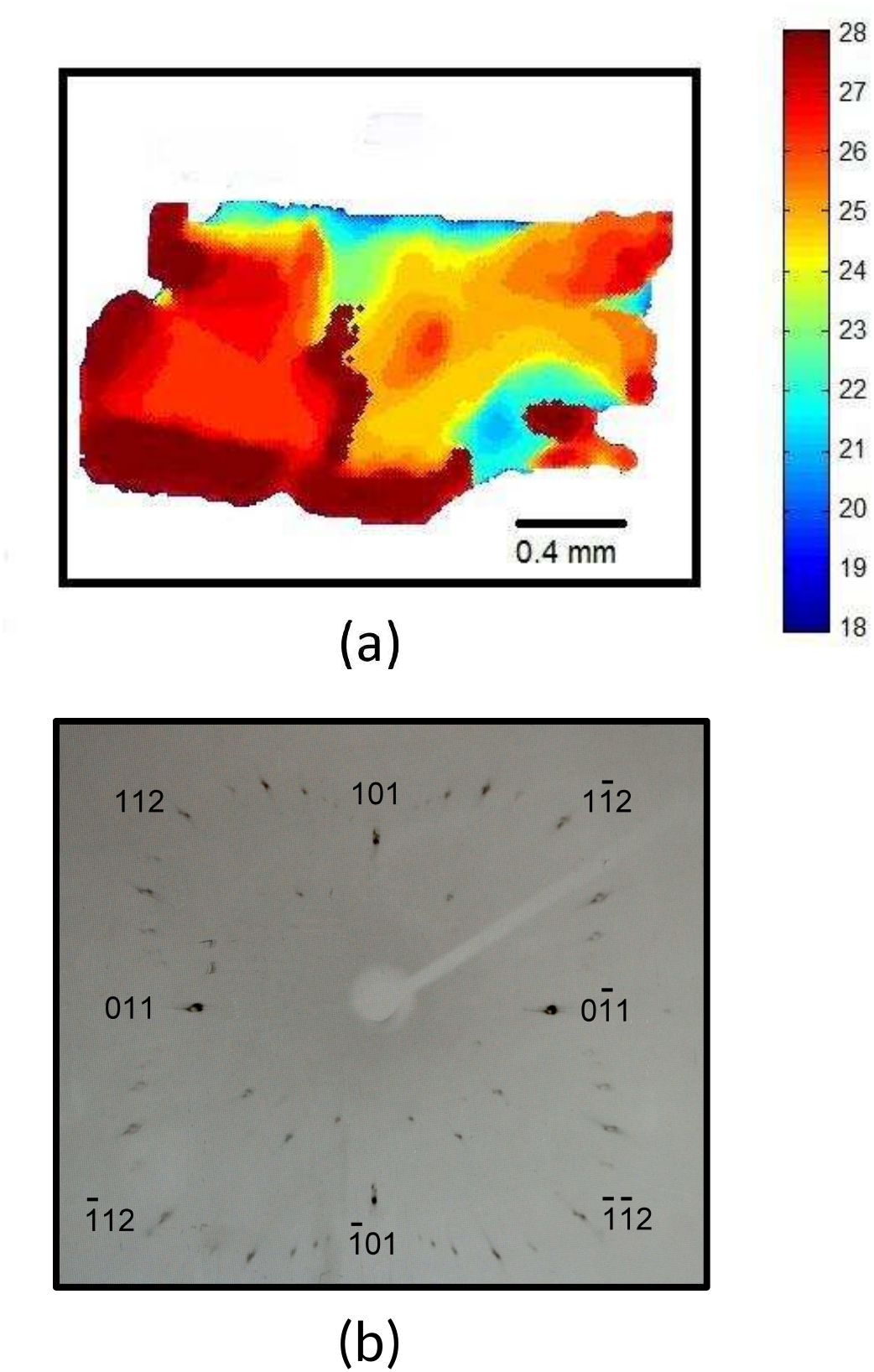}
\caption{\label{fig:Tcmap} (a) Surface map of the superconducting transition temperature ($T_c$) measured using a  scanning Hall probe magnetometer. The superconducting transition temperatures of the scanned surface are identified by color labels. (b) X-ray diffraction of the crystal measured using the transmission La{\"u}e method, showing the characteristic reflections of a tetragonal single crystal.}
\end{figure}

Figure \ref{fig:Tcmap}a shows the distribution of superconducting transition temperatures across the sample surface. A rather wide distribution of $T_c$ is observed, ranging from 22 to 28 K. However, a major part ($\sim$ 80$\%$) of the sample shows $T_c$~$\sim$~25-28~K. Our current sample is an interesting case in the sense that, in-spite of its somewhat wide $T_c$ distribution, the results show the presence of SMP and also a relatively high critical current density at low temperatures. It is also unusual that the SMP vanishes for temperatures above $T_c/2$. In this work, we studied this unexpected behavior of the SMP, using magnetic relaxation measurements. The single crystal was also analysed through X-ray diffraction using the transmission La{\"u}e method. Fig. \ref{fig:Tcmap}b shows the lauegram measured with the primary X-ray beam oriented perpendicular to the major face of the crystal. The distribution of reflections indicates that the direction normal to the larger plane of the sample corresponds to a quaternary axis of rotation (C4), which allows us to identify it with the direction of the c-axis of the tetragonal network of the compound. The spots have a slightly filamentous shape and were indexed from the cell parameters provided in Ref. [\cite{dir09}]. The x-ray diffraction results are characteristic of a single crystal material. The lauegram pattern was simulated by taking in to account the dispersion of cell parameters arising from a small variation of potassium concentration ($x$) within the sample, as inferred from the $T_c$ distribution in Fig. \ref{fig:Tcmap}a. The observed simulated reflections (not shown) suggest that the slightly filament-shaped spots might be related to crystalline domains in the sample with slightly different cell parameters, which in turn are associated with a small variation of potassium content within the sample. Overall, the x-ray results confirm the good crystalline character of the sample under study.

\begin{figure}
\centering
\includegraphics[height=6cm]{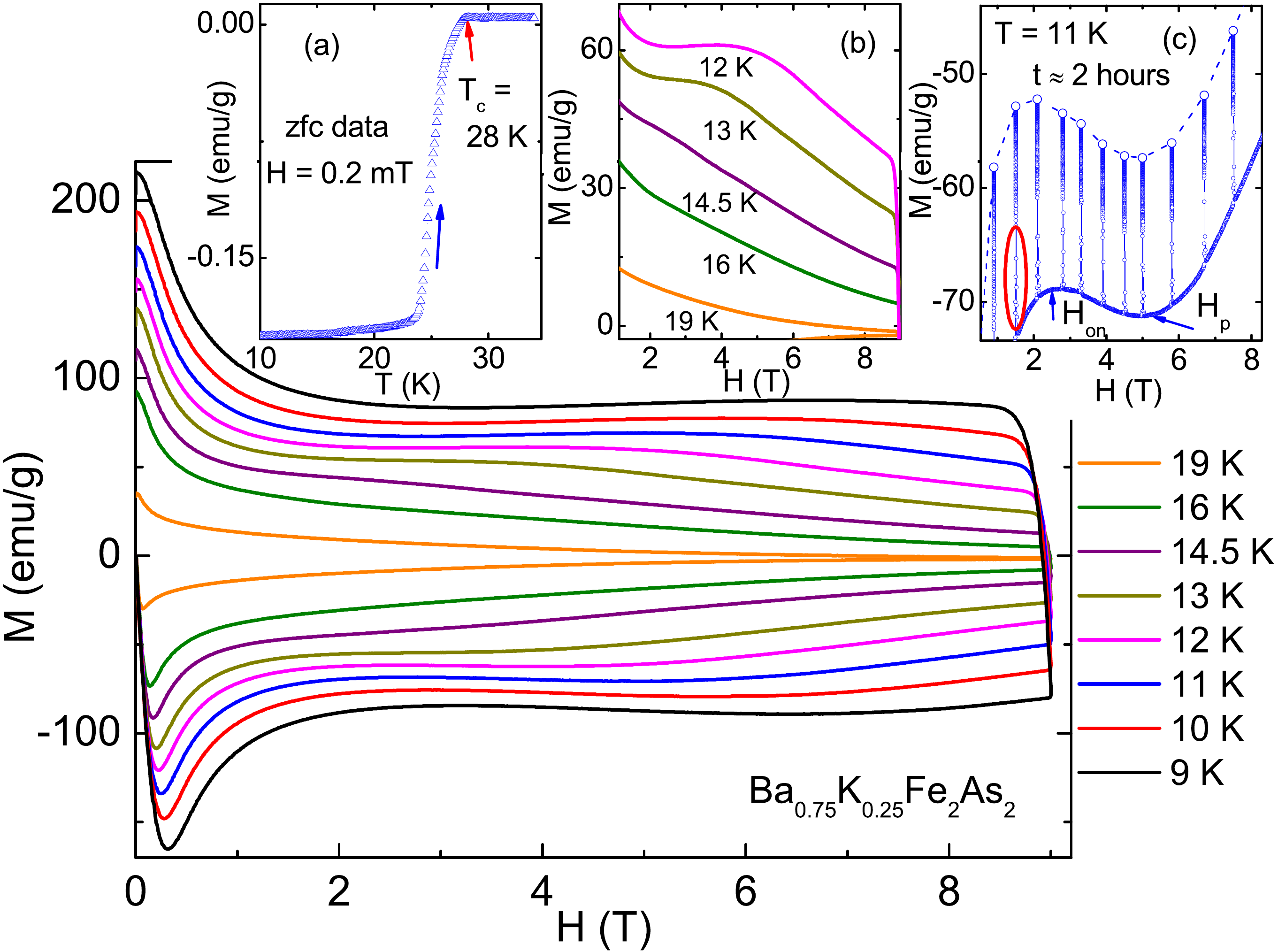}
\caption{\label{fig:MHandMT} Isothermal magnetic field dependence of magnetization, $M(H)$, for temperatures well below $T_c$, showing the signature of SMP. Inset ($a$): Temperature dependence of the zfc magnetization with $H$ = 0.2 mT, showing the onset of superconducting transition at 28 K. Inset ($b$): Selected isothermal $M(H)$ curves show a clear signature of SMP is observed only up to 12 K and vanished completely at temperatures higher than 14.5 K. Inset ($c$): Isothermal $M(H)$ at $T$ = 11 K with the magnetic relaxation data, measured for about 2 hours with different magnetic fields. The arrows indicate the $H_{on}$ and $H_p$ values as described in the text.}
\end{figure}

Figure \ref{fig:MHandMT} shows selected isothermal $M$($H$) curves evidencing the SMP appearing as the temperature decreases below 14.5 K. Figure \ref{fig:MHandMT}$a$ shows the transition temperature $T_c$. Figure \ref{fig:MHandMT}$b$ shows a detail of the upper branch of selected $M$($H$) curves evidencing that the SMP only develops below 14.5 K. Figure \ref{fig:MHandMT}$c$ shows magnetic relaxation data, $M$($time$), obtained for selected fields over the lower branch of isothermals $M$($H$) curves; at 11 K the onset field of the SMP, $H_{on}$, and the peak field, $H_p$, are well defined. Interestingly, the circle in this inset represents the first 30 seconds of relaxation, which corresponds to about 40$\%$ of the total magnetic relaxation in a 2 hour period. All magnetic relaxation curves showed the usual logarithmic behavior with time, $\mid$$M$$\mid$ $\sim$ log($t$) and plots of ln$\mid$$M$$\mid$ vs. ln$t$ allowed us to obtain the relaxation rate $R$ = dln$\mid$$M$$\mid$/dln$t$. 

\begin{figure}
\centering
\includegraphics[height=6cm]{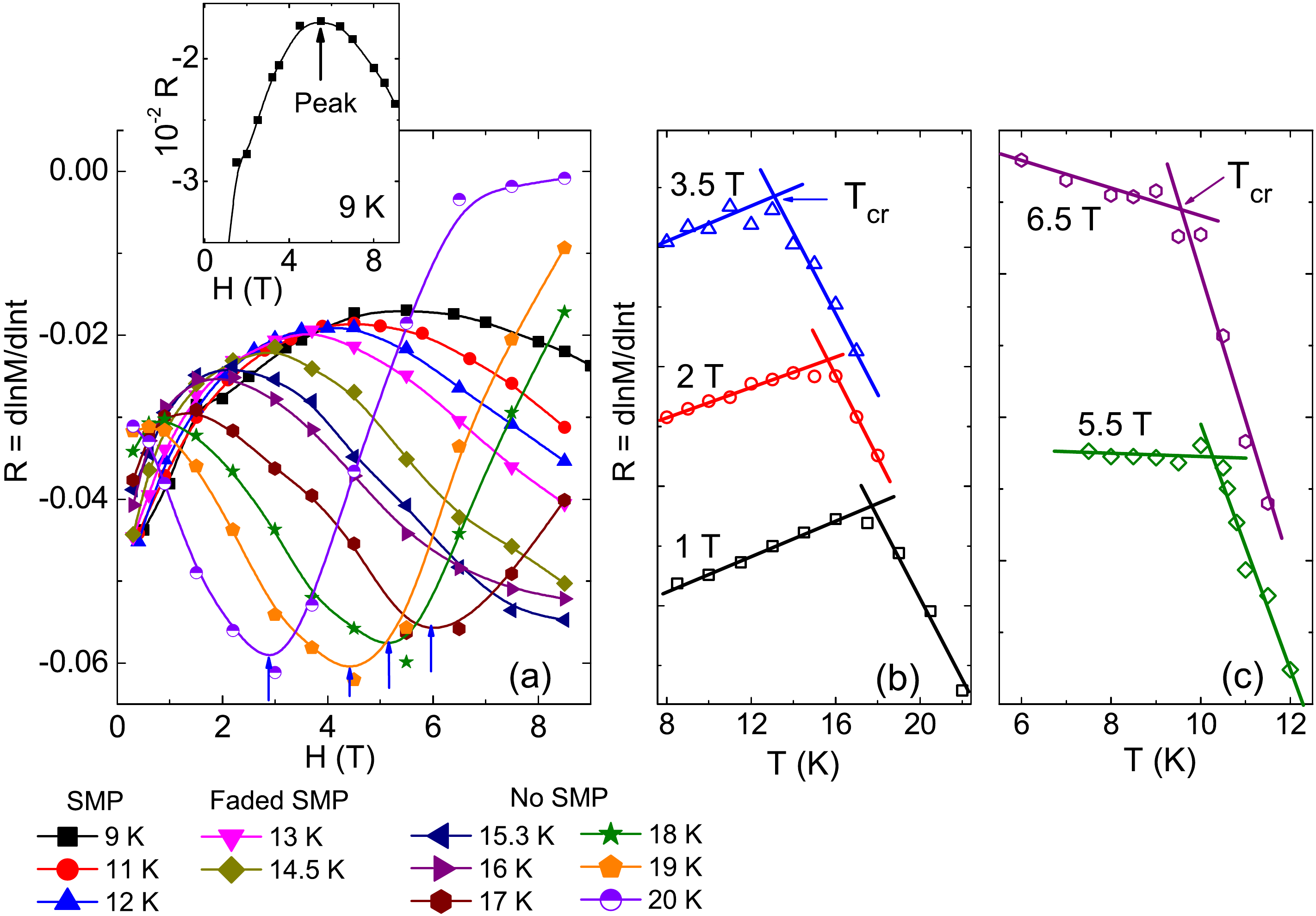}
\caption{\label{fig:relx_rate} (a): Magnetic field dependence of the relaxation rate, $R = dlnM/dlnt$, in the SMP, faded-SMP and no-SMP temperature regimes. In the SMP regime, $R$(H) shows the peak structure whereas in the no-SMP regime, $R$(H) shows the inverted peak. Inset shows the peak in $R$(H) at $T$ = 9 K.  (b) and (c): Temperature dependence of magnetic relaxation rate at different constant magnetic fields; in each panel, $T_{cr}$ represents the pinning crossover. In panels (b) and (c), the $R$ values lies in the range of 0.020 - 0.035, however, the curves are shifted upward for clarity.}
\end{figure}

Figure \ref{fig:relx_rate}$a$ shows plots of $R$ vs. $H$, where each curve represents values of $R$ obtained for a given isothermal $M$(H$)$. All $R$($H$) isothermal in Fig. \ref{fig:relx_rate}$a$ show a peak which shifts to lower fields as the temperature increases. While the peaks show some correspondence with $H_p$ in the respective $M$($H$) curves showing the SMP, the argument fails as there is no SMP above 14 K. The peak in each curve suggests a crossover from single vortex or collective pinning (depending on how high is the magnetic field in the region below the peak) to plastic pinning as plasticity is expected as $\mid$$R$$\mid$ increases. But as shown in Fig. \ref{fig:relx_rate}$a$, as $H$ increases an inverted peak appears for the higher temperature isothermals. Although the inverted peak would suggest some relation with $M$($H$) curves that do not show the SMP, it seems instead, that the effect is related to a proximity to the irreversible field, as probably the inverted peak would appear if a higher magnetic field were available for the lower temperature isothermals. Figures \ref{fig:relx_rate}$b$ and \ref{fig:relx_rate}$c$ show plots of the isofield $R$ vs. $T$ where two well defined different behaviors are observed evidencing a possible crossover at temperature $T_{cr}$ in the pinning mechanism. \cite{said15, kop10, pra11, miu12} It is interesting to observe that below $T_{cr}$ the slope of the $R(T)$ decreases slowly with increasing fields and shows only a change of slope and not a clear peak at higher fields (Fig. \ref{fig:relx_rate}$c$). As seen in Fig. \ref{fig:relx_rate}, the crossing points identified as $T_{cr}$ appear to be related to the position of $H_p$ in the $M$($H$) curves that show the SMP, but again the argument fails as there is no SMP in the isothermals above 14 K. These observations suggest that the vortex-dynamics in the different temperature regimes are different and we need a deeper insight to understand the suppressed SMP behaviour above 14 K ($T_c$/2).

\begin{figure}
\centering
\includegraphics[height=6cm]{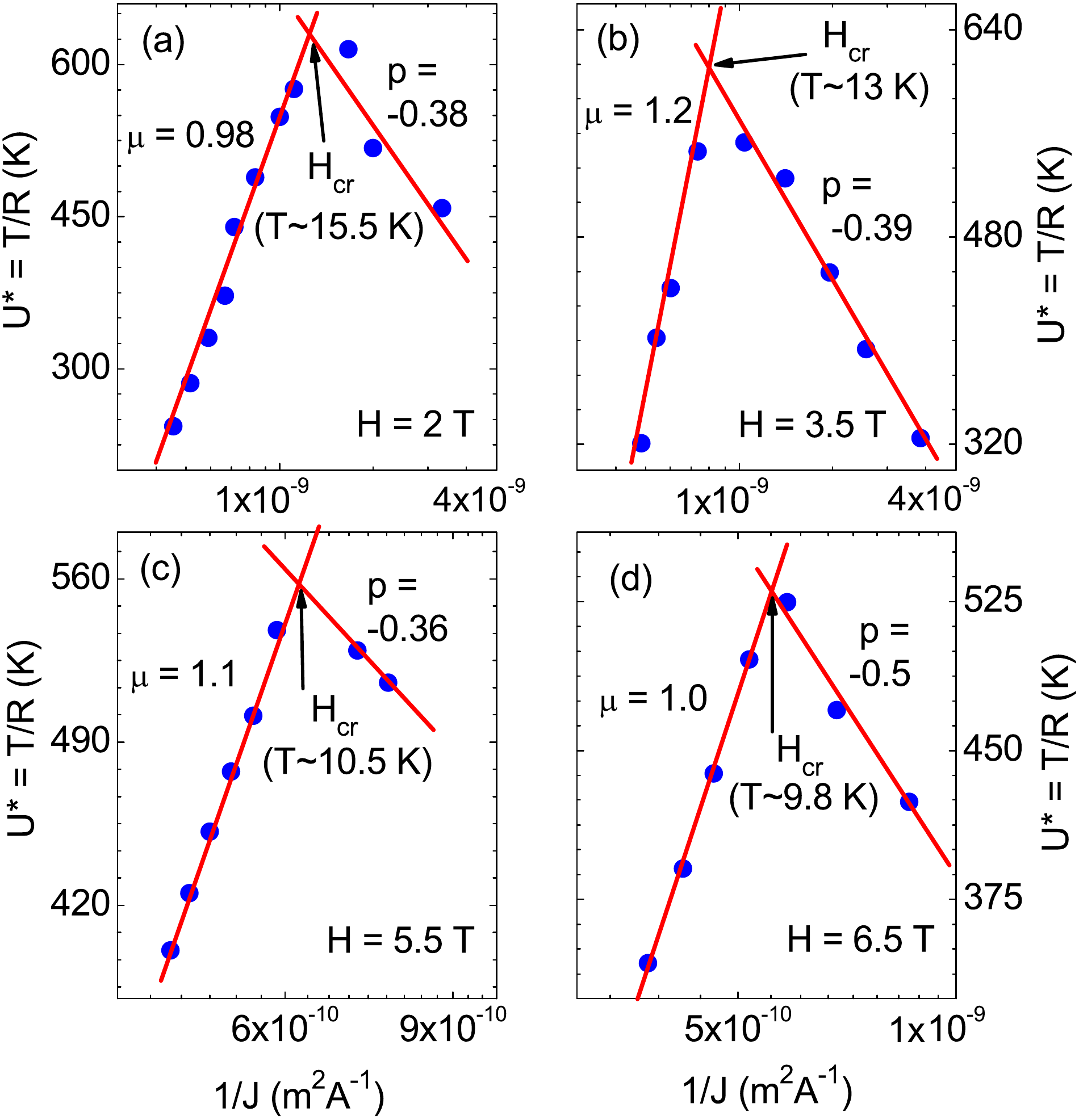}
\caption{\label{fig:Uvs1/J} Inverse current density (1/$J$) dependence of the activation energy ($U^*$). The arrows in each panel represents the pinning crossover between the different $U^*(1/J)$ dependencies.}
\end{figure}

\begin{figure}
\centering
\includegraphics[height=6cm]{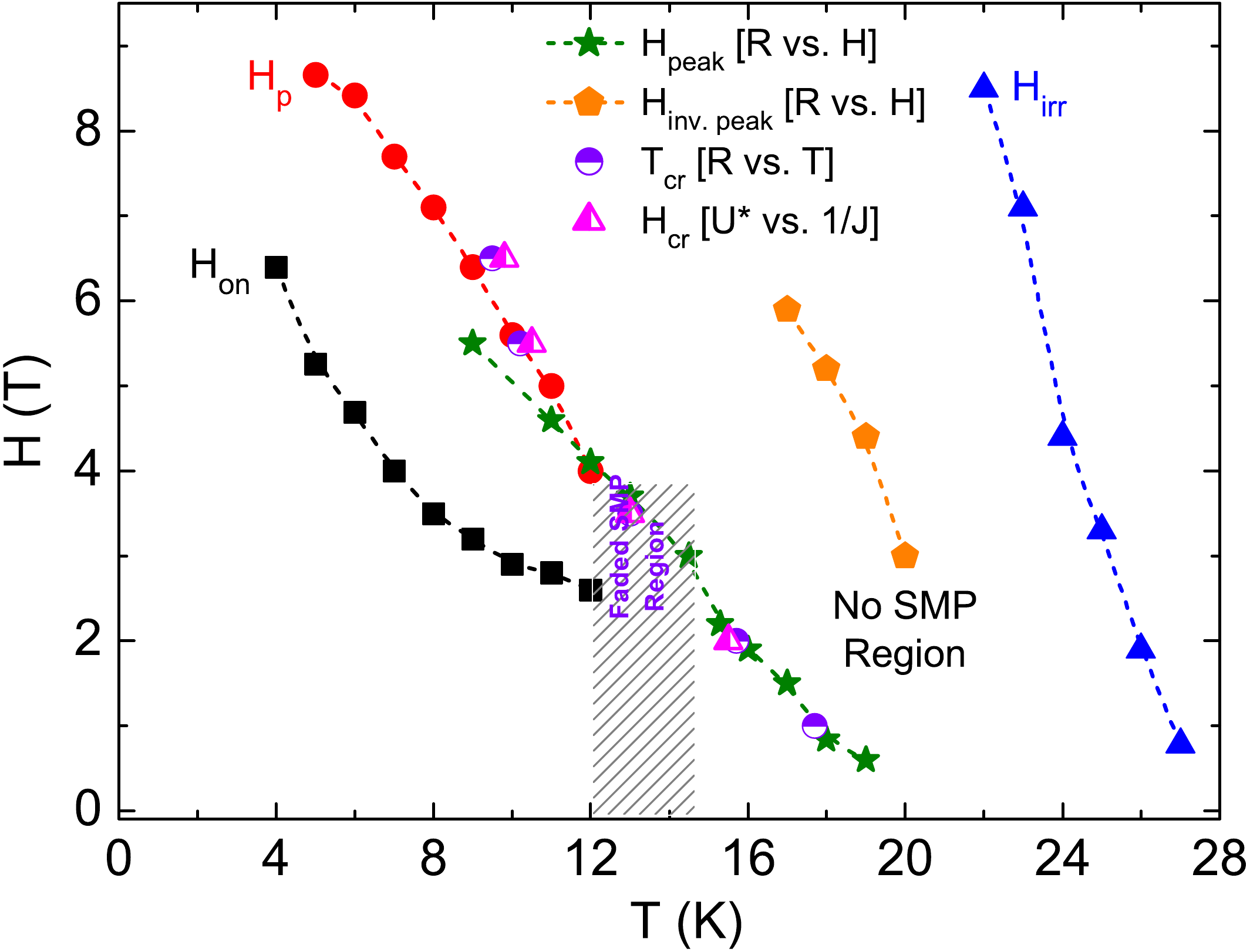}
\caption{\label{fig:HT_PD} $H$-$T$ phase diagram for the sample used in the present study. Different legends are explained in the text.}
\end{figure}

The $R(T)$ dependence discussed in the previous paragraph shows the validity of the collective pinning theory in the present case. Hence, it is useful to understand the vortex dynamics in the realm of collective pinning theory, using $U^* (= T/R)$ vs. $1/J$ plot, as previously exploited in recent studies. \cite{tos12, wei16} The activation energy $U^*$ in the collective pinning theory varies with the current density ($J$) as, $U^* = U_0 (J_c/J)^\mu$ \cite{fei89}, where $\mu$ and $J_c$ depends on the dimensionality and size of the vortex bundles under consideration. In the case of a 3-dimensional system, $\mu$ values were predicted to be 1/7, 3/2, 7/9 for single-vortex, small-bundle and large-bundle regimes, respectively. \cite{fei89, gri97} Hence, the exponent $\mu$ can easily be extracted by a double logarithmic plot of $U^*$ vs. $1/J$, shown in Fig. \ref{fig:Uvs1/J} for different fields. Our experiments yield $\mu$ values about $1.0-1.2$ at low temperatures (left side of $H_{cr}$), which lies in between 1/7 (single-vortex) and 3/2 (small-bundle). Similar $\mu$ values were reported in numerous other studies of IBS \cite{wei16, said10, miu12, hab11, tos12} and YBCO  superconductor, \cite{tho93} and suggest the different types of pinning contributions. On the other hand, the slope at high temperatures (right side of $H_{cr}$) is found to be about -$1/2$, which is consistent with the exponent observed in plastic-creep theory, \cite{abu96} where the negative exponent is usually denoted as $p$ with a value of -$1/2$. These observations suggest that there is a crossover from a collective pinning behaviour to plastic pinning, which gives rise to the SMP. However, the $H$-$T$ phase diagram (Fig. \ref{fig:HT_PD}) shows that the crossover point observed in $U^*$ vs. $1/J$ plot lies in the region where SMP doesn't exist. This discrepancy casts a shadow on the use of the analysis of $U^*$ vs. $1/J$ to show the collective to plastic-pinning crossover.

To analyse the behaviour of vortex-dynamics in different temperature regimes, we plotted the different characteristic temperature and field values in the phase diagram shown in Fig. \ref{fig:HT_PD}. Both $H_p$ and $H_{on}$ exist well below $H_{irr}$ line. The shaded portion in the diagram shows a region where SMP is not well defined, (named as faded-SMP), above which the typical signature of SMP in $M(H)$ vanishes (named as no-SMP). The $H_{peak}$ line (from $R(H)$) up to 12 K lies within the $H_p$ line, as also seen in Ref. \cite{wei16}, which shows that the SMP is associated with the peak in $R(H)$ in this temperature range. However, $H_{peak}$ is also present in the region above 12 K, where no SMP is observed. The $T_{cr}$ and $H_{cr}$ lines follow the $H_{peak}$ line in the faded-SMP and no-SMP regimes, whereas, in the SMP regime (below 12 K), it follow the $H_p$ line. This suggests that the crossover points $T_{cr}$ and $H_{cr}$ in $R$ vs. $T$ and $U^*$ vs. $1/J$ respectively are rather misleading in the faded-SMP and no-SMP temperature regimes. Hence, we employed another technique to investigate the vortex-dynamics in the SMP, faded-SMP and no-SMP temperature regimes, which is discussed below. 

\begin{figure}
\centering
\includegraphics[height=6cm]{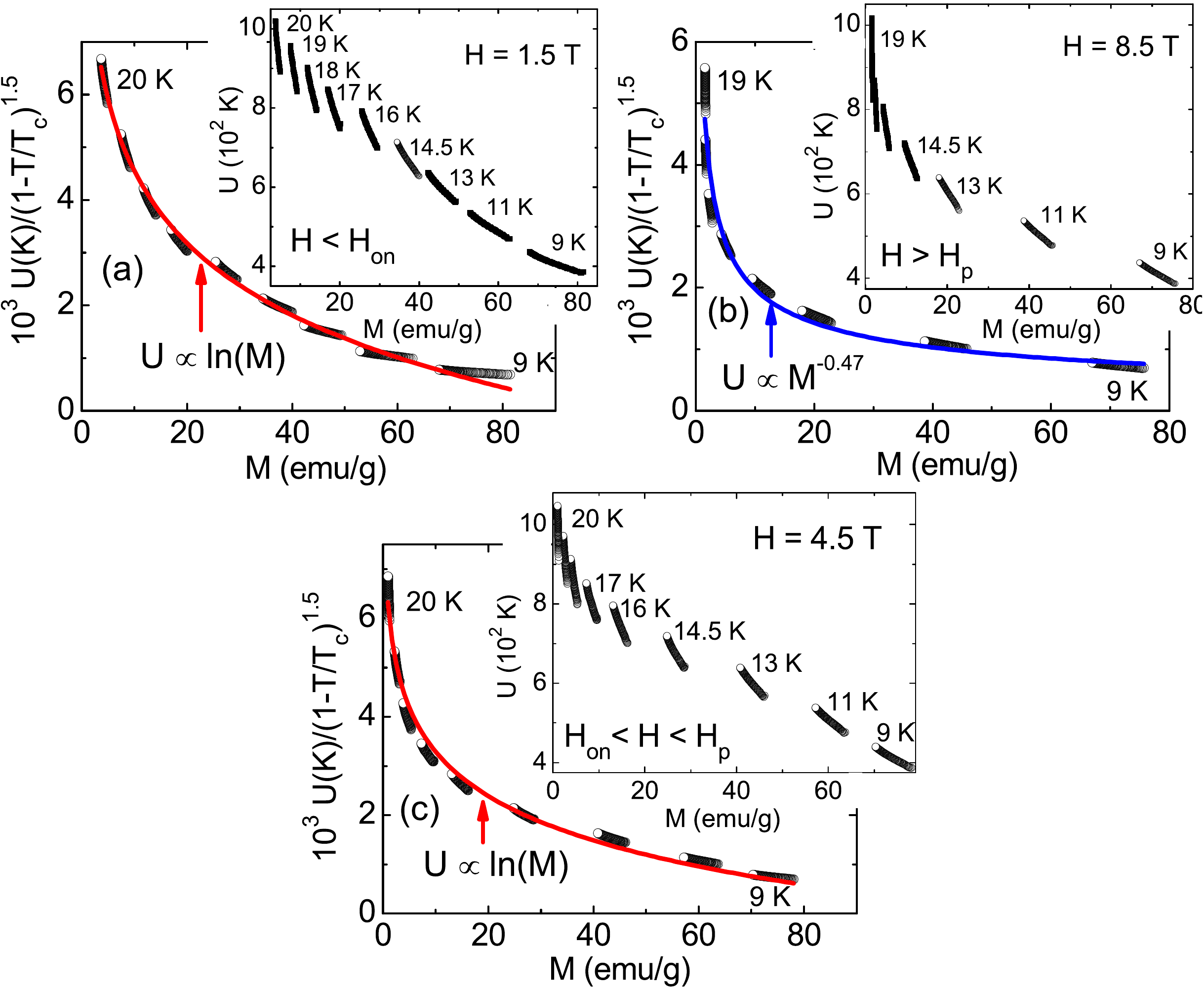}
\caption{\label{fig:maley_plot} Activation energy $U(M,T)$ at fixed fields, $H$ = 1.5 T, 4.5 T and 8.5 T after scaled using the function, $g(T/T_c) = (1-T/T_c)^{1.5}$. ($a$): for $H_{c1}$ $<$ $H$ $<$ $H_{on}$, ($b$): for $H$ $>$ $H_p$ and ($c$): for $H_{on}$ $<$ $H$ $<$ $H_p$. Inset of each panel shows $U(M)$  before scaling by $g(T/T_c)$ function.}
\end{figure}

To investigate the reason for the vanishing of the SMP above $T_c$/2, we studied the mechanism of vortex-dynamics using activation energy ($U$) vs. magnetic moment ($M$) curves. In this analysis, the activation energy $U(M)$ is obtained for each $M(t)$ curve by exploiting the approach developed by Maley et al., \cite{mal90} 
\begin{equation} \label{eq:1}
U = -Tln[dM(t)/dt] + CT
\end{equation}
where C is a constant which depends on the hoping distance of the vortex, the attempt frequency and the sample size. The activation energy is plotted with respect to the magnetic moment (obtained at fixed $H$ and $T$ during $M(t)$) in Fig. \ref{fig:maley_plot}. Panels ($a$), ($b$) and ($c$) display the $U$ vs. $M$ curves for $H$ = 1.5 T, 4.5 T and 8.5 T corresponding to the $H$$<$$H_{on}$, $H_{on}$$<$$H$$<$$H_p$ and $H$$>$$H_p$ respectively. The reason for choosing these field values for the Maley's plot is that the characteristic pinning mechanism in the field above and below the SMP might be different. The insets of Fig. \ref{fig:maley_plot}$a$, \ref{fig:maley_plot}$b$ and \ref{fig:maley_plot}$c$ show the results of Maley's analysis for $C$ = 40, which is justified below, similar C value has also been observed in the literature. \cite{hen91} In Fig. \ref{fig:maley_plot}, it is clear that the $U(M)$ curves do not show the smooth behaviour as has been observed for temperatures close to $T_c$. \cite{hen91} To obtain a smooth curve of $U(M)$, we have to scale the activation energy curves with $g(T/T_c)$ scaling function. \cite{hen91} As suggested by McHenry et al., \cite{hen91} in Fig. \ref{fig:maley_plot}, we have used the $g(T/T_c) = (1-T/T_c)^{1.5}$ scaling form to obtain the smooth $U(M)$ curves, which depends on the pinning length scale close to $T_c$.

The results of Fig. \ref{fig:maley_plot} show that the scaled $U(M)$ curve for $H$ = 4.5 T ($H_{on}$$<$$H$$<$$H_p$) follows a power law with $M^{-0.47}$. On the other hand, for fields $H$ = 1.5 T ($H_{c1}$ $<$ $H$ $<$ $H_{on}$) and 8.5 T ($H$ $>$ $H_p$), the scaled $U(M)$ curves follow a logarithmic behaviour. \cite{mal90, hen91} It is to be noted that the $C$ parameter is obtained to get a best smooth curve of $U(M)$. Using this analysis, we estimated $C$ = 40, which is then used to estimate the activation energy, $U(M)$, for each $M(t)$ curve in different temperature regimes (namely, SMP, faded-SMP, no-SMP). 

\begin{figure}
\centering
\includegraphics[height=6cm]{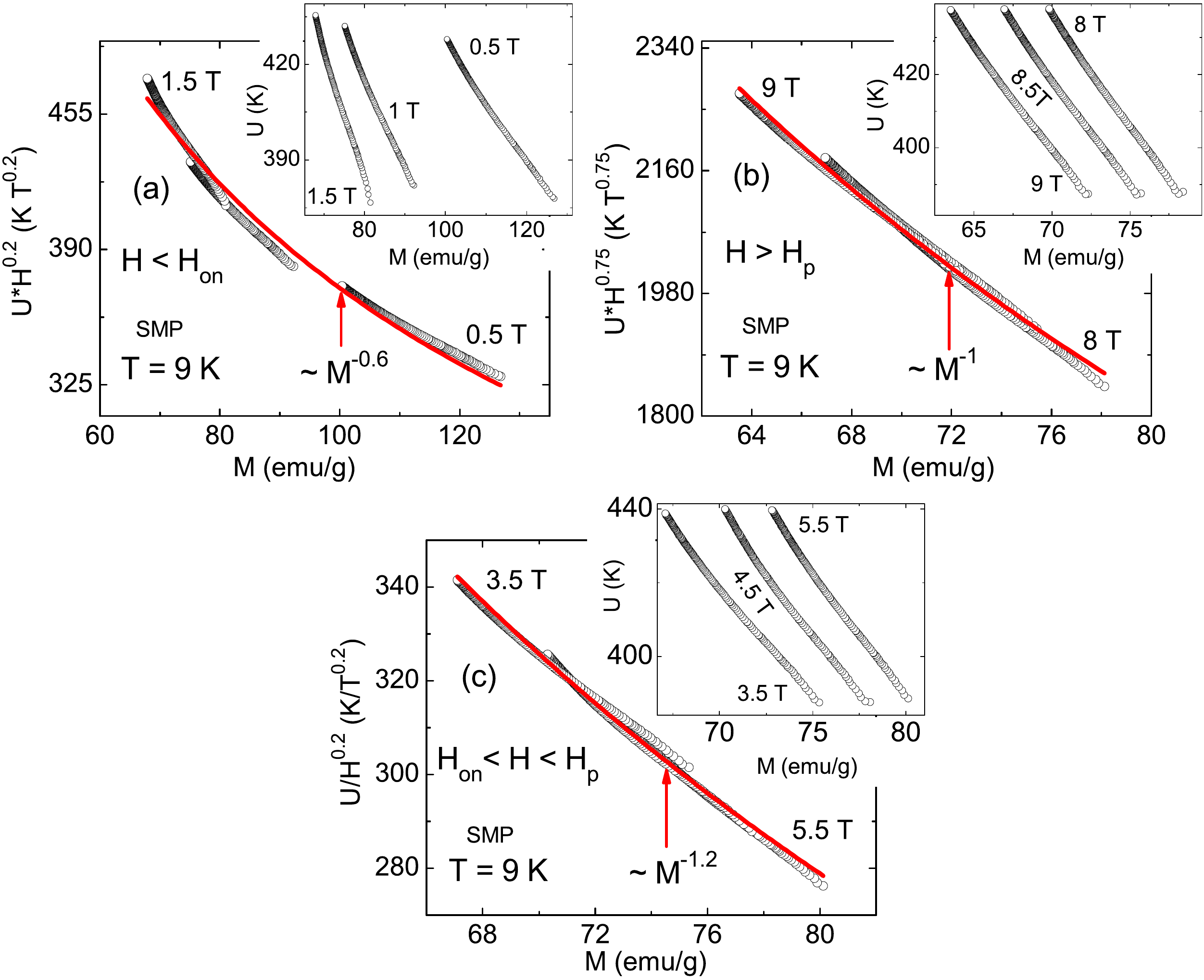}
\caption{\label{fig:PE_regime} The insets show $U(M)$ curves obtained through $M(t)$ measurements at $T$ = 9 K in different field regimes. Each main panel shows the respective scaled $U(M)$ curves with different exponents of $H$ in different field regimes. Each scaled curve shows the power law behaviour with $M$.}
\end{figure}

The insets of Fig. \ref{fig:PE_regime} show the $U(M)$ curves obtained at $T$ = 9 K for different field ranges. Since, the SMP is clearly observed at this temperature, we wish to investigate if the pinning mechanism is different for fields $H>Hp$ and $H<Hp$. As discussed previously \cite{fei89, bla94} and further exploited by Abulafia et. al., \cite{abu96} the expression for the activation energy in the collective creep theory is described as, $U(B,J) = B^{\nu}J^{-\mu}$ $\approx$ $H^{\nu}M^{-\mu}$, where the critical exponents $\nu$ and $\mu$ depend on the specific pinning regime. It is known that in the collective pinning, the activation energy increases with increasing magnetic field and later \cite{abu96} shown that the activation energy above $H_p$ is described in terms of plastic pinning, where the activation energy decreases with increasing field. It suggests that the positive or negative values of exponent $\nu$ defines the collective (elastic) and plastic pinning mechanisms respectively. With this understanding, we scaled the $U(M)$ curves shown in the insets for each panel of Fig. \ref{fig:PE_regime} for different magnetic field regimes at $T$ = 9 K with a different exponent of $H$. The scaled curves are shown in the main panels of Fig. \ref{fig:PE_regime} with smooth $U(M)$ behaviour. It is interesting to observe that each scaled curve follow a power law behaviour with $M$. Using Ref. \cite{abu96}, we may unambiguously state that in $H_{on}$$<$$H$$<$$H_p$, the vortex pinning is collective (Fig. \ref{fig:PE_regime}$c$) in nature and it changes to plastic pinning for $H$ $>$ $H_p$ (Fig. \ref{fig:PE_regime}$b$). Hence, the results of Fig. \ref{fig:PE_regime} clearly demonstrates that the SMP is due to a crossover from collective (elastic) to plastic pinning behaviour. 

On the other hand, the fitting of the scaled curve in Fig. \ref{fig:PE_regime}$a$, suggests that the vortex pinning for $H$ $<$ $H_{on}$ is also dominated by the plastic behaviour as for $H$ $>$ $H_p$. However, the activation energy increases with $H$ in the $H$ $<$ $H_{on}$ region, which is not consistent with the plastic behaviour of vortex-pinning. It is to be noted that the scaled $U(M)$ curve in Fig. \ref{fig:PE_regime}$a$ is obtained by considering that the $U(M)$ has a power law dependence with $H$ of the form $\approx H^{-0.2}$. This apparent contradiction in pinning behaviour for $H$ $<$ $H_{on}$ has been observed by Abulafia et al. \cite{abu96} and later by other researchers \cite{said10, wei16} and negates the possibility of plastic pinning for $H$ $<$ $H_{on}$. It also eliminates the possibility of collective pinning as observed for $H_{on}$$<$$H$$<$$H_p$. The nature of the pinning below $H_{on}$ may be understood in terms of the single vortex pinning, which changes to the collective pinning above $H_{on}$ and renders a peak at $H_{on}$. However, this peak is entirely different than the SMP observed at $H_p$ which arises due a pinning crossover (collective to plastic) below and above $H_p$  .    

\begin{figure}
\centering
\includegraphics[height=6cm]{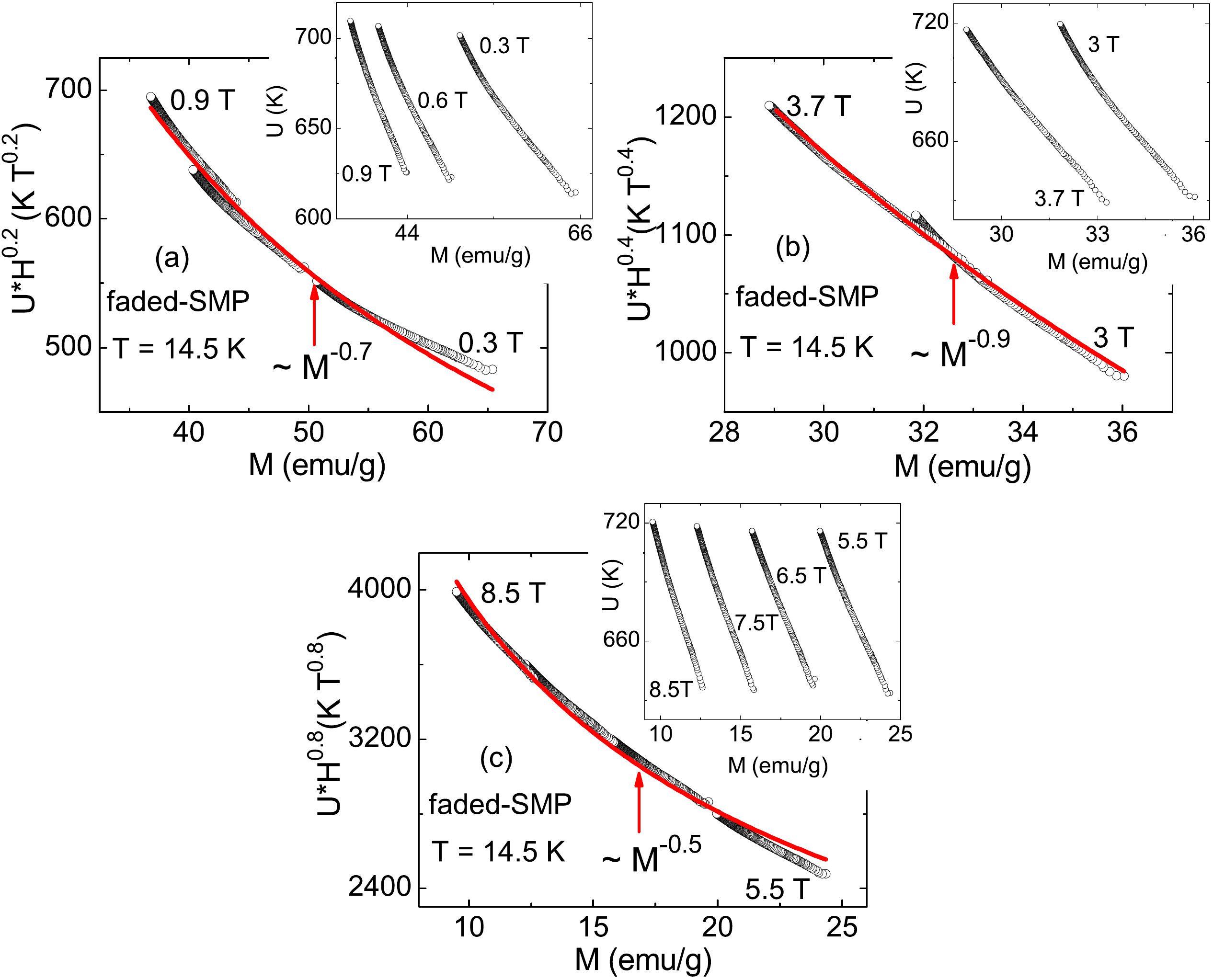}
\caption{\label{fig:fadedPE} The insets in each panel show $U(M)$ curves obtained through $M(t)$ measurements at $T$ = 14.5 K in different field regime. Each main panel shows the respective scaled $U(M)$ curve with an exponent of $H$ in different field regimes. Each scaled curve shows the power law behaviour with $M$.}
\end{figure}

Similarly to Fig. \ref{fig:PE_regime}, the activation energy $U(M)$ was also estimated from $M(t)$ curves measured at $T$ = 14.5 K, as shown in Fig. \ref{fig:fadedPE}. In Fig. \ref{fig:MHandMT} the $M(H)$ data at $T$ = 14.5 K shows a very subtle feature of SMP (faded-SMP). By comparing the pinning behavior at different field regimes shown in each panel of Fig.\ref{fig:fadedPE}, we can clearly state that the vortices show no pinning crossover (collective to plastic) as observed in Fig. \ref{fig:PE_regime}. This is an expected result for 14.5 K, where no distinct SMP is observed. The scaling shown in Fig. \ref{fig:fadedPE}$a$ points to a plastic pinning behaviour but in fact the plastic pinning is not valid in the mentioned field range and follows the single-vortex elastic pinning as also discussed in case of Fig. \ref{fig:PE_regime}$a$. Exactly the same results as at 14.5 K, are also observed for measurements at $T$ = 19~K  (not shown). Figure \ref{fig:fadedPE} emphasizes that the absence of SMP above $T_c/2$ is due to the absence of a pinning crossover (collective to plastic) at intermediate fields. This behaviour is somewhat unusual and to our knowledge has not been observed before. 

\begin{figure}
\centering
\includegraphics[height=7cm]{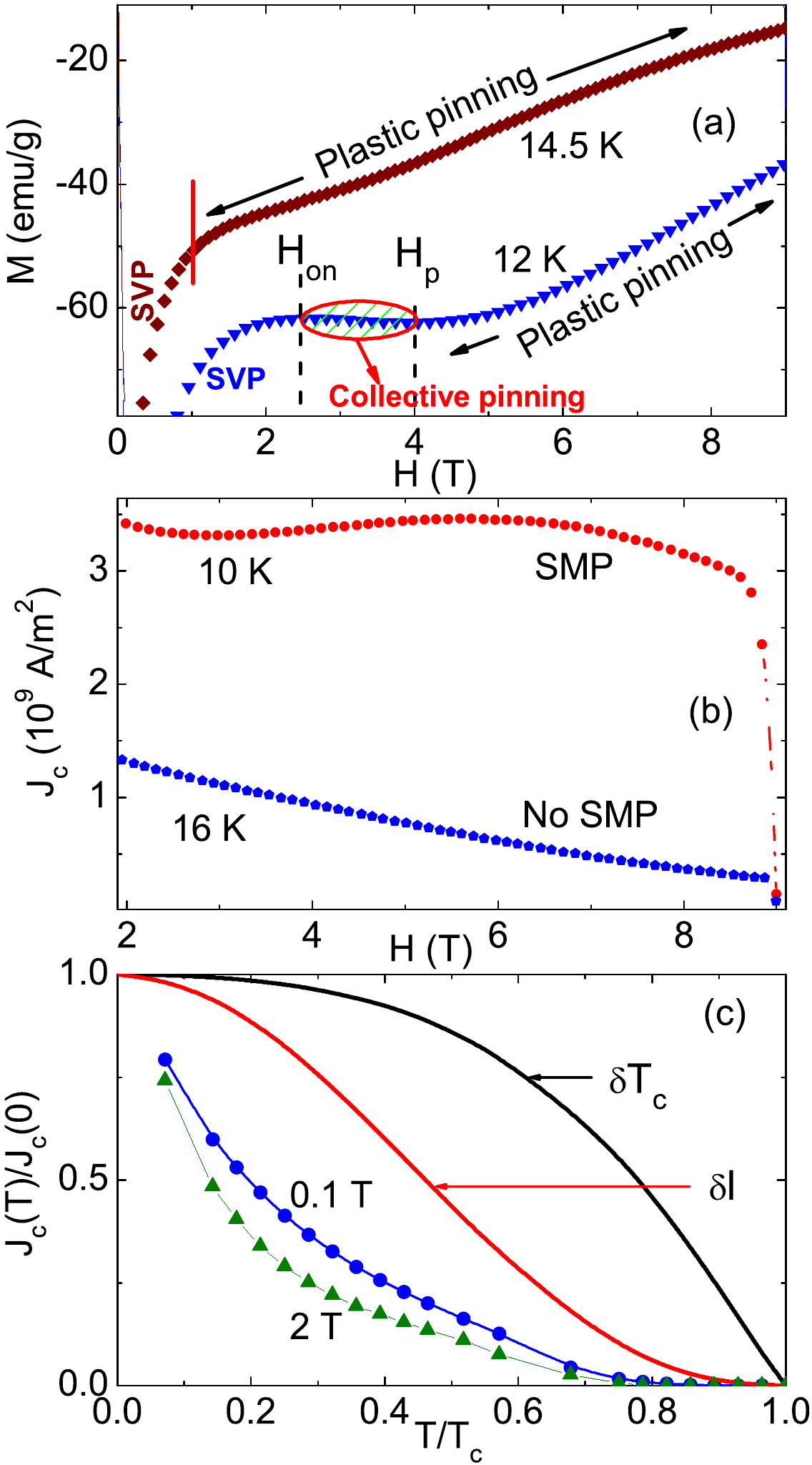}
\caption{\label{fig:Jc} (a) Magnetic field dependence of magnetization at 12 K and 14.5 K depict the vortex pinning behaviour in SMP and faded-SMP temperature regimes respectively. (b) Magnetic field dependence of critical current density at 10 K and 16 K corresponding to the SMP and no-SMP regimes respectively. (c) Normalized temperature ($T/T_c$) dependence of $J_c/J_c(0)$ at 0.1 T and 2 T magnetic fields. The solid lines represent the $\delta l$ and $\delta T_c$ pinning models.}
\end{figure}

Figure \ref{fig:Jc}$a$ shows $M(H)$ data, measured at 12 K and 14.5 K, to emphasize the difference in vortex creep and pinning in the SMP and faded-SMP temperature regimes respectively. In the SMP regime, a pinning crossover from collective to plastic creep is observed, whereas, in the faded-SMP regime, no such crossover is noticed, suggesting that plastic pinning takes the place of collective pinning. In the low field side, single vortex pinning (SVP) described the behavior of vortices. It must be noticed that the change in the pinning properties responsible for the disappearance of SMP above $T_c/2$, is not related with the sample inhomogeneity. Separate regions with different $T_C$ within the sample are distributed inhomogeneously, with a major part ($\sim$ 80$\%$) having $T_c$~$\sim$~25-28~K and a small part of the sample with a lower $T_c$ down to 22 K. Since the regions of different $T_c$ are spatially distributed quite separately from each other, it is unlikely that such inhomogeneity would contribute to the pinning mechanism of the sample. \cite{dew71} 
     
Figure \ref{fig:Jc}$b$ shows the magnetic field dependence of the critical current density, $J_c(H)$, at $T$ = 10 K and 16 K, estimated from the irreversibility in isothermal $M(H)$ above the field of full penetration. The $J_c$ was estimated with Bean's critical state model, \cite{bean64} using the expression $J_c = \Delta M[a_2(1-a_2/3a_1)]^{-1}$, where $\Delta M$ is the difference between the field decreasing and field increasing branches of $M$ (ref.\cite{ss15}). The parameters 2$a_1$ and 2$a_2$ are the dimensions of the sample to get the cross sectional area perpendicular to field direction. The $J_c$ value at $T$ = 10 K in the SMP regime is about $3.4 \times 10^9$ A/m$^2$ and is even higher at lower temperatures. High values of $J_c$ have been reported for YBCO tapes (epitaxial deposition and biaxially textured tapes), where the maximum $J_c$ is reported to be $7 \times 10^9$ A/m$^2$ in zero field at 77 K. \cite{goy96, dav96} In Co-doped BaFe$_2$As$_2$ iron-pnictide superconductors, the maximum $J_c$ value is found to be even higher than 10$^{10}$A/m$^2$ at 4.2 K for thin films. \cite{tak10, tak11a, tak11b} However, the polycrystalline bulk and round wires of Ba$_{0.6}$K$_{0.4}$Fe$_2$As$_2$ show a maximum $J_c$ of about 10$^9$A/m$^2$ at 4.2 K in self field, \cite{wei12} which is claimed to be more than 10 times higher than that of any other round untextured ferropnictide wire and even 4-5 times higher than the best textured flat wire. \cite{wei12} Therefore, the $J_c$ value in the present study ($3.4 \times 10^9$ A/m$^2$) is comparable as found by the Weiss et al. \cite{wei12} for optimally doped BaKFe$_2$As$_2$ superconductor. This suggests that an underdoped BaKFe$_2$As$_2$ superconductor is also a potential candidate for application. These results are consistent with the recent observations by Song et. al. \cite{son16} 

Figure \ref{fig:Jc}$c$ shows the reduced temperature ($T/T_c=t$) dependence of $J_c(T)/J_c(0)$ at 0.1 T and 2 T. This figure clearly shows that the pinning mechanism in the present sample is much different than the conventional $\delta l$ ($J_c(t)/J_c(0)=(1-t^2)^{5/2}(1+t^2)^{-1/2}$) and $\delta T_c$ ($J_c(t)/J_c(0)=(1-t^2)^{7/6}(1+t)^{5/6}$) pinning models, which has been observed in the other IBS of 122 family. \cite{vla15} It has also been argued that this unconventional pinning behaviour in the 122 family is directly related with the strong intrinsic pinning and need further attention. \cite{vla15} However, the literature \cite{gho12} suggests that, in Ba$_{0.72}$K$_{0.28}$Fe$_2$As$_2$ superconductor, both the $\delta l$ and $\delta T_c$ pinning co-exist and their contributions are strongly temperature and magnetic field dependent.

\section{Conclusions}
In conclusion, we presented a study of isofield magnetic relaxation and isothermal magnetization on an underdoped Ba$_{1-x}$K$_{x}$Fe$_2$As$_2$ ($x$ = 0.25) single crystal, with a superconducting transition temperature $T_c$ = 28 K. The sample exhibits the second magnetization peak (SMP) for temperatures below $T_c/2$ and this feature is suppressed completely at higher temperatures. The temperature regime below 12 K is defined as the SMP regime and above 14.5 K, as the no-SMP regime, whereas, at intermediate temperatures, we call it the faded-SMP regime. The gradual suppression of the SMP above 12 K has been studied initially through the field dependence of relaxation rate, $R(H)$. It shows a peak behavior at the low field side with an inverted peak at higher fields and both of these peaks shift to the lower field side at higher temperatures. The peak in $R(H)$ shows good resemblance with the observed position ($H_p$) of the SMP and extends up to the no-SMP temperature region. The crossover point ($H_{cr}$) in the plot of the inverse current density (1/$J$) vs. activation energy ($U^*$)  also follows the peak position in $R(H)$ in the no-SMP region, and in the SMP region, it follows the SMP line. The crossover point $H_{cr}$ suggests the elastic to plastic pinning crossover even in the no-SMP region. To resolve this apparent controversy, relaxation data, $M(t)$, was used to study the vortex-dynamics in different regimes, using Maley's approach. The results suggest that the observed SMP below 12~K is due to collective to plastic pinning crossover whereas in the temperature regime above 12 K, no pinning crossover has been observed and plastic pinning replaces the collective pinning behavior. In addition, the critical current density, $J_c$ was estimated using Bean's model and found to be about $3.4 \times 10^9$ A/m$^2$ at T = 10~K (in SMP regime), which is comparable to the value observed for optimally doped BaKFe$_2$As$_2$ superconductor. The pinning mechanism in the present case is much different than the conventional $\delta l$ and $\delta T_c$ pinning models, and, is similar to the other IBS of 122 family. The higher value of $J_c$ in this underdoped pnictide suggests its usefulness for technological application over optimally doped samples.
 \section{ACKNOWLEDGMENTS}
This work was supported by the Brazillian agencies CAPES (Science without Borders program), FAPERJ and CNPq.

\end{document}